# Spectroscopic study of unique line broadening and inversion in low pressure microwave generated water plasmas

R. L. Mills,[a)] P. C. Ray, R. M. Mayo, M. Nansteel, B. Dhandapani

*BlackLight Power, Inc., 493 Old Trenton Road, Cranbury, NJ 08512*

Jonathan Phillips

*University of New Mexico, Dept. of Chemical and Nuclear Engineering, 203 Farris Engineering, Albuquerque, NM 87131*

It was demonstrated that low pressure ($\sim 0.2$ Torr) water vapor plasmas generated in a 10 mm inner diameter quartz tube with an Evenson microwave cavity show at least two features which are not explained by conventional plasma models. First, significant (> 2.5 Å) hydrogen Balmer $\alpha$ line broadening, of constant width, up to 5 cm from the microwave coupler was recorded. Only hydrogen, and not oxygen, showed significant line broadening. This feature, observed previously in hydrogen-containing mixed gas plasmas generated with high voltage dc and rf discharges was explained by some researchers to result from acceleration of hydrogen ions near the cathode. This explanation cannot apply to the line broadening observed in the (electrodeless) microwave plasmas generated in this work, particularly at distances as great as 5 cm from the microwave coupler. Second, inversion of the line intensities of both the Lyman and Balmer series, again, at distances up to 5 cm from the coupler, were observed. The line inversion suggests the existence of a hitherto unknown source of pumping of the optical power in plasmas. Finally, it is notable that other aspects of the plasma including the OH$^*$ rotational temperature and low electron concentrations are quite typical of plasmas of this type.

a) Author to whom correspondence should be addressed; Phone: 609-490-1090; Fax: 609-490-1066; Electronic mail: rmills@blacklightpower.com

## I. INTRODUCTION

About a decade ago scientists began to study the emission spectroscopy of "mixed gas" plasmas in detail, and these studies revealed some surprising results for those mixed gas plasmas in which hydrogen was one of the gases. First, in mixtures of argon and hydrogen, the hydrogen emission lines are significantly broader than any argon line.[1–14] Second, many such plasmas show a peculiar non-Boltzmann population of the excited states. Both the Lyman and the Balmer $\alpha$ lines are "overpopulated."[8,10,11,15,16]

We recently reported a new and related phenomenon for mixed gas plasmas containing hydrogen: Preferential hydrogen line broadening, often extreme, was found in a number of mixed gas discharge plasmas. Specifically, it has been reported that in several mixed gas systems Balmer $\alpha$ lines broader than 2.5 Å were created in He/$H_2$ (10%) and Ar/$H_2$ (10%) as well as plasmas containing volatilized Sr such as Sr/$H_2$, Sr/Ar/$H_2$ (10%) and Sr/He/$H_2$ (10%) plasmas.[8,9,11,13,14] In none of these plasmas was there any significant broadening of the noble gas lines. Furthermore, hydrogen lines were not broadened in a number of other mixed gas plasmas including Xe/$H_2$ (10%), and Kr/$H_2$ (10%). These results show that the presence of hydrogen line broadening in mixed gas plasmas is not limited to Ar/$H_2$, and they also suggest, consistent with the Mills' model of atomic hydrogen,[11,17–23] that only very special and predictable systems will show this broadening.

Related work,[19–23] shows there are other unique spectral features to those mixed gas plasmas containing hydrogen in which selective hydrogen line broadening is found. Specifically, unique vacuum ultraviolet (VUV) emission lines are found at wavelengths precisely predicted by a new theory (outlined in numerous publication including reference 19–23). Specifically strong lines were found at wavelengths corresponding to energies of q*13.6 where q=1,2,3,4,6,7,8,9,11. These strong emissions are not found in any single gas plasma, and cannot be assigned to the known emission of any of the single gases studied. These lines are only found in mixed gas plasmas, called resonance transfer (RT) plasmas. Indeed, those mixed gas plasmas (*e.g.* Kr/$H_2$ and Xe/$H_2$) that do not produce line broadening, also do not produce any special "mixed gas plasma" lines in the VUV.[19–23]

In the present paper we report some remarkable features of a water vapor (mixed gas) plasma generated using an Evenson microwave cavity. (As both oxygen and hydrogen are quickly evolved in the plasma this can be considered a "mixed gas" plasma containing hydrogen.) First, it was found that low pressure (<1 Torr) water vapor plasmas generated with the Evenson microwave cavity, have preferential hydrogen line broadening similar to that described in earlier reports on the mixed gas plasmas. Second, line broadening is found to persist, undiminished, up to 5 cm from the coupler. In addition, evidence of strong pumping to cause population inversion is presented. At distances up to several centimeters from the cavity, inversion (negative temperatures in a Boltzmann distribution) were recorded for the Balmer series. Calculations indicate that the degree of inversion observed may require as much as 200 W · $cm^{-3}$ pumping power in at least a limited volume of the plasma.[24,25] At a microwave input power of 9 W · $cm^{-3}$, a collisional radiative model showed that the hydrogen excited state population distribution was consistent with an *n* = 1→5,6 pumping power of 200 W · $cm^{-3}$, a far higher pumping power density than that found in existing laser systems. Yet, other measurements show quite ordinary features. For example, the rotation temperature profile determined from $OH^*$ species was not unusual in any aspect. The concentration of charged species was found to be very low. It is also notable that there was no evidence of pumping or line broadening when a radio frequency coil replaced the microwave coupler on our system.

## II. EXPERIMENTAL

The plasma reactors employed in this work were simple quartz tubes, 1 cm diameter by 20 cm in length. All of the visible spectroscopy was collected with the cell in the configuration shown Fig. 1. The microwave cavity used was an E-mode[26,27] co-axial Evenson cavity microwave (Opthos Model #B1, powered with 2.45 GHz Opthos MPG–4M generator) which only "covers" a section of the reactor about 2.5 cm in length. For all microwave studies net input power was maintained at 90 W. Reflected power was kept to less than 5 W. For the 0.2 Torr operation mass loss studies showed the flow rate of water vapor to be approximately 1 sccm.

For the rf plasma study, the only change was that the microwave cavity was removed and replaced with a coil, about 10 cm in total length, wrapped around the central section of the tube. The rf coil was powered from a 13.56 MHz generator (RF VII, Inc., Model MN 500) with a matching network (RF Power Products, Inc., Model RF 5S, 300 W). The coil inductance and resistance were 4.7 $\mu$H and 0.106 $\Omega$, respectively. The coil impedance was 400 $\Omega$ at 13.56 MHz. The forward rf power was 90 W, and the reflected power was less than 1 W. The gas flow rates and pressure were 1 sccm and 0.2 Torr, respectively.

For the VUV study of a water plasma, a similar quartz tube was employed with the same Evenson cavity, but the tube end opposite the water source was opened to the VUV spectrometer through a pin-hole collimator. This configuration was employed as UV "transparent" optics do not exist. Using differential pumping (more detail, reference 19) it was possible to operate the water plasma for the VUV study at 0.2 Torr, while the spectrometer itself was operated at about $10^{-6}$ Torr.

The high resolution visible spectrometer (scanned between 4000–7000 Å using a 0.05 Å step size) was a Jobin Yvon Horiba 1250 M with 2400 groves/mm ion-etched holographic diffraction grating with a resolution of ± 0.06 Å over the spectral range 1900–8600 Å. The entrance and exit slits were set to 20 $\mu$m. The signal was recorded by a photomultiplier tube (PMT) with a stand alone high voltage power supply (950 V) and an acquisition controller. The data was obtained in a single accumulation with a 1 second integration time. The plasma emission was fiber-optically coupled through a 220F matching fiber adapter positioned 2 cm from the cell wall to the spectrometer. Using this geometry, we were able to show with a red laser that light was collected from a source length in the cell of no greater than 2 mm. Thus, there was no "overlap" in the "light source" for each of the data points collected.

The response of this spectrometer was shown to be approximately flat in the 4000–7000 Å region with an intensity calibrated tungsten lamp. To measure the absolute intensity, the spectrometer and detection system were calibrated[28] with 5460.8 Å, 5769.6 Å, and 6965.4 Å light from a Hg–Ar lamp (Ocean Optics, model HG-1) that was calibrated with a NIST certified silicon photodiode. The population density of the $n = 3$ hydrogen excited state $N_3$ was determined from the absolute intensity of the Balmer $\alpha$ (6562.8 Å) line measured using the calibrated spectrometer. In turn, the absolute intensities of $n = 4$ to 9 were determined from the ratio of each of these lines to that of the absolute intensity Balmer $\alpha$ ($n = 3$). Additional detail on this spectrometer and its operation are given elsewhere.[9,24,25]

The VUV spectrometer (20–5600Å) was a normal incidence 0.2 meter monochromator equipped with a 1200 lines/mm holographic grating with a platinum coating. The VUV spectrum was recorded with a channel electron multiplier (CEM). The wavelength resolution was about 0.2 Å full width half maximum (FWHM) with slit widths of 50 $\mu$m. The increment was 2 Å and the dwell time was 500 ms. The VUV spectra (900–1300 Å) of the water and control hydrogen plasmas were recorded at 90 W input power.

The spectrometer was calibrated between 400–2000 Å with a standard discharge light source using He, Ne, Ar, Kr, and Xe lines: He I (584 Å), He II (304 Å), Ne I (735 Å), Ne II (460.7 Å), Ar I (1048 Å), Ar II (932 Å), Kr II (964 Å), Xe I (1295.6 Å), Xe II (1041.3 Å), Xe II (1100.43 Å). The wavelength and intensity ratios matched those given by NIST.[29] The spectrometer response was determined to be approximately flat in the 1000–1300 Å region.

Measurements (upper bound) of the electron density and temperature of the water plasma were determined using a compensated Langmuir probe according to the method given previously.[30]

## III. RESULTS

In order to definitively demonstrate the unique behavior of water vapor plasmas generated with an Evenson cavity, we thoroughly characterized both an Evenson cavity microwave plasma and an rf plasma maintained in the exact same tube operated at the same pressure and approximate flow rate. As the following data shows, the Evenson cavity generated plasma showed remarkable features, including excessive hydrogen atom line broadening and excited state population inversion; whereas, the spectroscopic study of the rf plasma indicated conventional behavior with no inversion. Comparing these results allows us to demonstrate that our measurement technique is sound and that we were able to measure features with high fidelity.

For water vapor plasmas, the following parameters were measured as a function of distance from the center of the microwave coupler: (i) the hydrogen Balmer $\alpha$ line width, (ii) the gas temperature determined using $OH^*$ rotation, (iii) line intensities expressing the excited state population inversion, and (iv) tube surface temperature profiles measured with a series of type-K thermocouples. An attempt was made to directly measure and spatially map the electron density in the microwave plasma using a Langmuir probe. However, the recorded signals were below the detection limit indicating $n_e < 10^9$ cm$^{-3}$ $n_e < 10^9$ cm$^{-3}$.

The rotational temperature profile (based on the population of OH$^*$ levels), of the microwave and rf plasmas are shown in Fig. 2. It is clear that the rotational temperature profile (Fig. 2) is qualitatively similar for all the plasmas and is similar to that of any single gas plasma. The gas temperature profile, determined from the OH$^*$ band using the Boltzmann plot method as described in detail elsewhere,[31] declines with distance from the plasma coupler. Also, the temperatures are clearly in the range anticipated given the results of earlier studies of gas temperatures in microwave plasmas[31–35] over a range of operating conditions.

In sharp contrast to the "standard" behavior of rotation temperature profiles, is the observation that there is an inversion of the excitation state populations. Indeed, the relative intensities of the Balmer series of lines fitted using the standard Boltzmann approach yield "negative temperatures." The generation of a non-Boltzmann distribution of electron populations in a gas or solid system results from some mechanism of pumping.[36,37] As noted later for inversion and excessive H line broadening as well, it cannot simply reflect energy transfer from hot electrons. Thus, this finding implies that there is some form of energy transfer in this "mixed gas" plasma that cannot be ascribed to the standard model of microwave plasmas whereby electrons acquire electric field energy and transfer it to ions and neutrals only thorough collisions. Furthermore, inversion is observed at distances as great as 5 cm from the coupler. The relative intensity spectra collected in several cases are shown in Fig. 3, and plots of relative line intensities for these "inverted" populations are shown in Fig. 4.

The measured hydrogen line broadening as a function of position is shown in Fig. 5 as FWHM of the H$_\alpha$ (6562.8 Å) line versus distance along the tube reactor from the zero position at the center of the Evenson cavity. Once again, it was found that in a mixed plasma only the hydrogen line is broadened, and it is broadened significantly. A new finding is that the line broadening is axially uniform in the plasma tube up to 5 cm from the coupler rather than rapidly diminishing with distance from the power source as in the case with excessively broadened H in Ar/H$_2$ rf and glow discharges.[1–7] This new finding is also not consistent with any model that attributes selective line broadening to either a field (extremely low outside the coupler region) or high electron densities (very low in this region).

It should also be noted that at distances greater than about 5 cm beyond the coupler neither inversion nor line broadening was observed. Even at this position, the relative line intensities were not in proportion to that expected from a thermal or Boltzmann distribution as the Balmer $\beta$ state was overpopulated.

We were also able to collect limited data regarding the Lyman series of lines using the VUV spectrometer tuned to shorter wavelengths. Unfortunately, this system is not readily adaptable for mapping the plasma as a function of position. Only a "global" spectrum, that is a sampling of light from the entire plasma, is possible (Fig. 6). Still, it is clear in the figure that for the plasma as a whole, the excited state population inversion impacts the Lyman series as well. This provides a verification that a source of pumping impacts the hydrogen atom excited state populations.

The truly unique nature of the 0.2 Torr water vapor plasma created with the Evenson cavity can best be understood by comparison with other plasmas. For example, running the system at a slightly higher pressure reduced the extent of the inversion significantly. As shown in Figs. 3 and 4, the inversion extended less than 3 cm from the microwave coupler at this pressure.

Another comparison was made using the same system (quartz tube, water supply, pump), but replacing the Evenson cavity with an rf coil. This generated a discharge with none of the unique features of the microwave plasma. That is, as shown, for an rf source run with approximately the same power as the Evenson cavity there is no excited state population inversion (Figs. 3 and 4), or line broadening (Fig. 5). Yet, the rotation temperature profile of the two plasmas is very similar (Fig. 2). Figure 4 provides a direct comparison of the relative populations of excited states for the two systems. Clearly for the rf plasma the states are populated as anticipated for a Boltzmann distribution; whereas, the Evenson cavity excited population is "exceptional." That is, the Evenson cavity results indicate a source of pumping; whereas, there is no pumping for the rf system.

One additional observation of importance to evaluating the model of Mills ('RT plasma', references 19–23) for providing the high energies required for the selective line broadening is the existence of significant quantities of $O^{2+}$ at 3715.0 Å, 3754.8 Å, and 3791.28 Å as shown in Fig. 7. Again, these lines, like all non H-atom lines in the microwave generated water plasma, show no broadening.

## IV. DISCUSSION

Consistent with earlier work, the data collected for this study clearly show that in "mixed gas" plasmas containing hydrogen, the hydrogen lines are extraordinary broadened, indicating that hydrogen atoms are 10–100 times more energetic than other molecular and atomic species present. In addition, we report a unique feature of one particular low pressure, hydrogen containing mixed gas plasma. Specifically, low pressure (∼0.2 Torr) water vapor plasmas generated with an Evenson cavity show a significant inversion of excited state populations. In contrast, a water vapor rf plasma created with the same nominal input power, the same tube, and at the same pressure, showed no unusual features. Indeed, not even broadening of the hydrogen lines was observed. Moreover; both plasmas showed "expected" behavior in terms of rotation temperature and very low electron densities. This data alone pose some puzzling questions.

What plasma mechanism could provide the significant power density required to "pump" electrons into the non-thermal distribution of excited states? Why are the unusual features only found for the Evenson cavity plasma?

We have assumed that Doppler broadening due to thermal motion was the dominant source to the extent that other sources may be neglected. This assumption was confirmed when each source was considered. In general, the experimental profile is a convolution of two Doppler profiles, an instrumental profile, the natural (lifetime) profile, Stark profiles, van der Waals profiles, a resonance profile, and fine structure. The contribution from each source was determined to be below the limit of detection.[8,9,14] This same conclusion was reached by other workers detecting selective H-atom line broadening in mixed gas plasmas.[1–14]

Mapping a number of characteristics of the plasma as a function of linear distance from the microwave coupler raises additional questions. First, the finding that the line broadening persists, undiminished, up to 5 cm from the coupler, provides evidence that earlier models of the line broadening mechanism do not apply to this system. In those models hydrogen line broadening is attributed to hydrogen ion acceleration by a static electric field. In the present system, there can be no acceleration in the region outside the cavity region as there is no field. Also, there is no known mechanism for significant ion acceleration in a microwave field, indicating there is no mechanism of field acceleration in the cavity. Moreover, even if there was a hitherto unknown process of direct ion acceleration in the cavity, this effect should be dramatically diminished by collisions long before an ion travels 5 cm. Indeed, the mean free path for "hard sphere" collisions is less than about 1 mm for this system. Yet, experimental measures clearly show no change in line shape even 5 cm from the coupler.

Furthermore, only the electrons respond to the microwave field rather than hydrogen species, $H^+, H_2^+, H_3^+, H^-$, H, or $H_2$. But, the measured electron temperature was about 1 eV;

whereas, the measured H temperature was 50 eV. This requires that $T_H >>> T_e$. This result also can not be explained by electric field acceleration of charged species. Nor can it be explained by electron or external Stark broadening. The electron density of about $10^8$ cm$^{-3}$ was five orders of magnitude too low.[8,9,14] And, in microwave driven plasmas, there is no high electric field in a cathode fall region (> 1 kV/cm) to accelerate positive ions as proposed previously[1-7] to explain significant broadening in hydrogen containing plasmas driven at a high voltage electrodes. Thus, it is impossible for H or any H-containing ion which may give rise to H to have a higher temperature than the electrons in a microwave plasma.

In sum, conventional arguments would suggest that the excess energy responsible for the observed hydrogen line broadening would be dissipated through collisions within a couple of millimeters of its "generation." Since the lifetimes of H ($n = 2$) and H ($n = 3$) are each approximately $10^{-8}$ s and the average velocity of the fastest hydrogen atoms was $<10^5$ m/s, the excitation must have been local.[5] In turn, this indicates that the process creating the line broadening is not restricted to the coupler zone, and thus must be occurring under conditions not dependent on direct input of microwave power.

Another difficulty for conventional theory is to explain the relationship between the H line broadening and non-Boltzmann populations of the excited states. Hydrogen line broadening is unique to certain plasmas such as argon and helium plasmas mixed with hydrogen[8,9,11-14] as well as water vapor plasmas. Inversion is unique to water plasmas, but H line intensity profile anomalies are also observed for the former plasmas as well. Mills *et al.* reported previously[8] that upon the addition of 5% argon to a hydrogen microwave plasma, the Lyman $\alpha$ emission was observed to increase by about an order of magnitude; whereas, xenon control had no effect. A similar result was obtained in the case of hollow-cathode hydrogen glow discharge plasmas to which 5% argon or helium was added.[10,11] Anomalous Ar/H$_2$ plasmas were observed by other researchers. Fozza *et. al.*[15] recorded the EUV spectrum of a $H_2$ microwave plasma and observed that the ratio of the intensity of the Lyman $\alpha$ emission to the molecular hydrogen emission (Lyman $B^1\Sigma_u^+ \to X^1\Sigma_g^+$) was about two orders of magnitude in the case of gas flow ratio of $\phi_{H_2/Ar} : \phi_{Ar} = 1 : 14$ versus the pure hydrogen plasma at twice the input power. Furthermore, Fujimoto *et al.*[38] have determined the cross section for production of excited hydrogen atoms from the emission cross sections for Lyman and Balmer lines when molecular hydrogen is dissociated into excited atoms by electron collisions. This data was used to develop a collisional-radiative model to be used in determining the ratio of molecular-to-atomic hydrogen densities in tokomak plasmas. Their results indicate an excitation threshold of 17 eV for Lyman $\alpha$ emission. Addition of other gases would be expected to decrease the intensity of hydrogen lines which could be absorbed by the gas. Hollander and Wertheimer[16] found that within a selected range of parameters, a helium-hydrogen microwave plasma emits a very intense hydrogen Lyman $\alpha$ radiation at 1215 Å which is up to 40 times more intense than other lines in the spectrum. The Lyman $\alpha$ emission intensity showed a significant deviation from that predicted by the model of Fujimoto *et al.*[38] and from the emission of hydrogen alone. An internal source of free energy could explain the line broadening and non-Boltzmann populations of the excited states observed previously,[8–14] and further studies with controls[19–23] could confirm the source in independent studies.[15,16]

The conditions of an extraordinarily high H Doppler temperature, a negative H excitation temperature, and cold electron, excitation, and rotational temperatures measured on OH* is impossible to explain with collisional-radiative, coronal, or Saha-Boltzmann models.[39] We did an exhaustive search for a possible conventional mechanism that was consistent with all of the results. Whatever explanation is adopted, it should explain several phenomena simultaneously. Particularly, the mechanism must account for the extraordinary line broadening up to 6 cm from the coupler in a region devoid of electrons and must simultaneously explain the inversion of excited state populations. It should further explain why these phenomena are plasma-source dependent (*i.e.* why the phenomena occur in an Evenson microwave plasma, but not in an inductively coupled rf plasma maintained in the same plasma cell). Not one of the unique features of the observed H-atom spectra could be independently explained satisfactorily using standard models. The Frank–Condon effect (FCE) was even considered. This effect creates H with energies of about 2 eV by wall recombination reactions of the type[40,41]

$$H_2^+ + e^- \rightarrow 2H \qquad (1)$$

In our experiments, Evenson microwave water plasmas were used as sources of O, $O_2$ and atomic hydrogen. As only the Evenson cavity appears to generate the unique phenomena observed here, this suggests a linkage with the special characteristics of plasmas created with this device. Earlier literature indicates that the charge species concentration in such a plasma is quite low.[8,9,14,24] That was observed in this system as well. Moreover, the concentration of species such as H, O, and $O_2$ are expected to be very high. The latter condition is required for the Mills' model of unique "reactions" involving hydrogen and various catalytic species (O-atoms and molecules in this system), a model which also postulates that line broadening processes are not strong "field" related, and hence not restricted to the coupler region. In this case, the pumping power source giving rise to inversion is the same as the broadening power source.[24,25,42,43]

## V. CONCLUSION

In conclusion, there is a growing body of evidence that mixed gas plasmas of particular compositions produce emission spectra that cannot be readily explained by any classical model of plasmas. That is, there is now a set of observations for mixed gas plasmas, some made in multiple labs, including extreme line broadening of only one species (H), the demonstration of pumping of population inversion in H, and the repeated discovery of unique spectral lines in the VUV region whose energies are integer-squared those of the energies of H, which challenge traditional theory. Finally, it must be noted that the evidence is consistent with a novel energy transfer mechanism.

REFERENCES

[1] M. Kuraica and N. Konjevic, Phys. Rev. A **46**, 4429 (1992).

[2] M. Kuraica, N. Konjevic, M. Platisa, and D. Pantelic, Spectrochim. Acta **47**, 1173 (1992).

[3] I. R. Videnovic, N. Konjevic, and M. M. Kuraica, Spectrochim. Acta, Part B **51**, 1707 (1996).

[4] S. Alexiou and E. Leboucher-Dalimier, Phys. Rev. E **60**, 3436 (1999).

[5] S. Djurovic and J. R. Roberts, J. Appl. Phys. **74**, 6558 (1993).

[6] S. B. Radovanov, K. Dzierzega, J. R. Roberts, and J. K. Olthoff, Appl. Phys. Lett. **66**, 2637 (1995).

[7] S. B. Radovanov, J. K. Olthoff, R. J. Van Brunt, and S. Djurovic, J. Appl. Phys. **78**, 746 (1995).

[8] R. L. Mills and P. Ray, New J. Phys. **4,** 22.1 (2002).

[9] R. L. Mills, P. Ray, B. Dhandapani, R. M. Mayo and J. He. J. Appl. Phys. **92**, 7008 (2002).

[10] R. Mills, J. Dong, W. Good, P. Ray, J. He, and B. Dhandapani, Int. J. Hydrogen Energy **27**, 967 (2002).

[11] R. Mills, M. Nansteel, and P. Ray, IEEE Trans. Plasma Sci. **30,** 639 (2002).

[12] R. Mills, M. Nansteel, and P. Ray, J. Plasma Phys. **69**, 131 (2003).

[13] R. Mills, M. Nansteel, and P. Ray, New J. Phys. **4**, 70.1 (2002).

[14] R. Mills, P. Ray, B. Dhandapani, and J. He, IEEE Trans. Plasma Sci. **31** (3), 338 (2003).

[15] A. C. Fozza, M. Moisan, M. R. Wertheimer, J. Appl. Phys. **88**, 20 (2000).

[16] A. Hollander, and M. R. Wertheimer, J. Vac. Sci. Technol., A **12**, 879 (1994).

[17] R. Mills, Int. J. Hydrogen Energy **27**, 565 (2002).

[18] R. Mills, J. Dong, and Y. Lu, Int. J. Hydrogen Energy, **25**, 919 (2000).

[19] R. Mills and P. Ray, J. Phys. D: Appl. Phys. **36**, 1535 (2003).

[20] R. L. Mills, P. Ray, B. Dhandapani, M. Nansteel, X. Chen, and J. He, J Mol. Struct. **643**, 43 (2002).

[21] R. Mills and P. Ray, Int. J. Hydrogen Energy **27**, 301 (2002)

[22] R. Mills and P. Ray, Int. J. Hydrogen Energy **27**, 533 (2002).

[23] R. L. Mills, P. Ray, J. Dong, M. Nansteel, B. Dhandapani, and J. He, Vib. Spectrosc. **31**, 195 (2003).

[24] R. Mills, P. Ray, and R. M. Mayo, submitted, available at http://www.blacklightpower.com/pdf/technical/H2Olaser%20paper8_09_02.pdf

[25] R. Mills, P. Ray, and R. Mayo, Appl. Phys. Lett. **82**(11), 1679 (2003).

[26] F. C. Fehsenfeld, K. M. Evenson, and H. P. Broida, Rev. Sci. Instrum. **35**, 294 (1965).

[27] B. McCarroll, Rev. Sci. Instrum. **41**, 279 (1970).

[28] J. Tadic, I. Juranic, and G. K. Moortgat, J. Photochem. Photobiol., A **143**, 169 (2000).

[29] NIST Atomic Spectra Database, www.physics.nist.gov/cgi-bin/AtData/display.ksh.

[30] F. F. Chen, "Electric Probes" in *Plasma Diagnostic Techniques*, R. H. Huddleston and S. L. Leonard, Eds. (Academic Press, NY, 1965).

[31] C. K. Chen and J. Phillips, J. Phys. D: Appl. Phys. **35**, 998 (2002).

[32] N. H. Bings, M. Olschewski, and J.A.C. Broekaert, Spectrochim. Acta **52B**, 1965 (1997).

[33] K. Tanabe, H. Haraguchi, and K. Fuwa, Spectrochim. Acta, **38B**, 49 (1983).

[34] J. Jonkers, J. M. de Regt, J. A. M. van der Mullen, H. P. C. Vos, F. P. J. de Groote, and E. A. H. Timmermans, Spectrochim. Acta **51B**, 1385 (1996).

[35] E. A. H. Timmermans, J. Jonkers, I. A. J. Thomas, A. Rodero, M. C. Quintero, A. Sola, A. Gamero, and J. A. M. van der Mullen, Spectrochim. Acta **53B**, 1553 (1998).

[36] J. J. Ewing, "Excimer Lasers" in *Laser Handbook*, M. L. Stitch, ed. (North-Holland Publishing Company, Amsterdam, 1979), Vol. A4.

[37] F. T. Arecchi and E. O. Schulz-Dubois, eds, *Laser Handbook* (North-Holland Publishing Company, Amsterdam, 1972), Vol. 1-6.

[38] T. Fujimoto, K. Sawada, and K. Takahata, J. Appl. Phys. **66**, 2315 (1989).

[39] H. R. Griem, *Spectral Line Broadening in Plasmas* (Academic Press, NY, 1978).

[40] D. R. Sweetman, Phys. Rev. Let. **3**, 425 (1959).

[41] D. R. Sweetman, Proc. Phys. Soc. Lon. **78**, 1215 (1961), Proc. R. Soc. Lon. A **256**, 416 (1960).

[42] R. Mills, P. Ray, and R. M. Mayo, IEEE Trans. Plasma Sci. **31** (2), 236 (2003).

[43] R. Mills and P. Ray, J. Phys. D: Appl. Phys. **36**, 1504 (2003).

FIGURE CAPTIONS

FIG. 1: Schematic of the Water Plasma System. Note, the only modification made for the rf study was the replacement of the Evenson microwave applicator and generator with an rf coil and generator.

FIG. 2: Gas Temperature Profiles. Rotational temperature ($T_R$) as a function of distance along the tube reactor from the zero position at the center of the Evenson cavity for three water plasmas. The Evenson cavity was operated at 0.2 Torr (a), and 1 Torr (b), and the rf system was operated at 0.2 Torr (c).

FIG. 3: Excited State Population Inversion. The visible emission spectrum (2700–7200 Å) from an Evenson water plasma operated at 0.2 Torr clearly shows excited state population inversion of the hydrogen atoms at 1 cm (a) and 5 cm (b) from the coupler, as well as at all points in between. Only at 7 cm from the coupler (c) is the population of the excited states in the expected proportion for a thermal plasma. The Evenson cavity operated at 1 Torr pressure doesn't show inversion at 4 cm (d) from the cavity, but does at 1 cm (e). In contrast, the rf plasma generated at 0.2 Torr doesn't show inversion at any position including 1 cm (f) and 5 cm (g) from the center.

FIG. 4: Relative Balmer Series Intensities. The relative intensities of Balmer series transitions ($I_n' = \frac{I_n}{Ag}$ where $I_n$ is the absolute intensity of level $n$, $A$ is the Einstein coefficient, and $g$ is the statistical weight given by $2n^2$) collected at 1 cm from the coupler center are shown for the Evenson cavity plasma operated at 0.2 Torr (a), 1.0 Torr (b), and for the rf plasma operated at 0.2 Torr (c).

FIG. 5: Line Broadening Profiles. The extraordinary line width for the H Balmer $\alpha$ line (6562.8 Å) emitted from the water plasma was found to vary little as a function of position from the center of the Evenson cavity operated at 0.2 Torr (a) and 1.0 Torr (b). Very little line broadening was found for the rf water plasma operated at 0.2 Torr (c). No line broadening was found for the oxygen radical (3971 Å) line for either the Evenson (d) or rf (e) water plasmas operated at 0.2 Torr.

FIG. 6: Lyman Series Populations. A VUV spectroscopic study of the Lyman series in a 0.2 Torr water plasma generated with an Evenson cavity clearly shows the impact of population inversion of the excited states on the relative intensities of the lines (a). In contrast, a pure hydrogen plasma at the same pressure shows no inversion (b).

FIG. 7: O (III) Emission. The visible spectrum (3700–3960 Å) of the cell emission from a water microwave plasma with 90 W input power. The catalysis mechanism was supported by the observation of $O^{2+}$ at 3715.0 Å, 3754.8 Å, and 3791.28 Å. $O^+$ was observed at 3727.2 Å, 3749.4 Å, 3771 Å, 3872 Å, and 3946.3 Å. The hydrogen Balmer lines corresponding to the transitions 10-2, 9-2, and 8-2 were also observed.

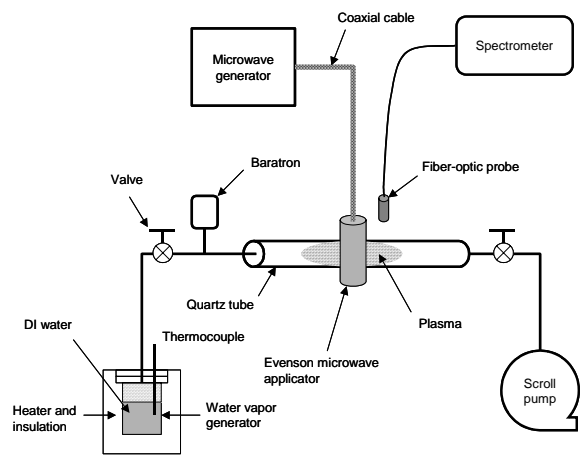

Figure 1

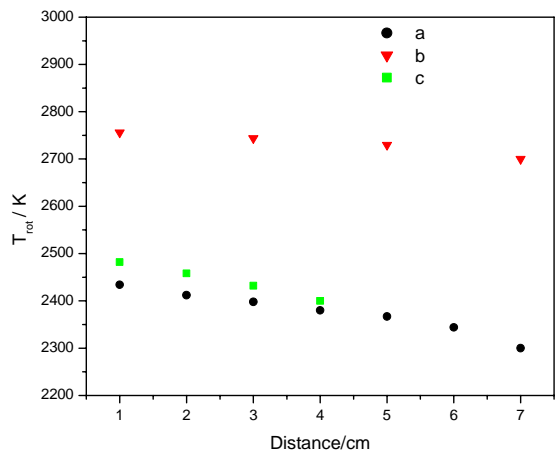

Figure 2

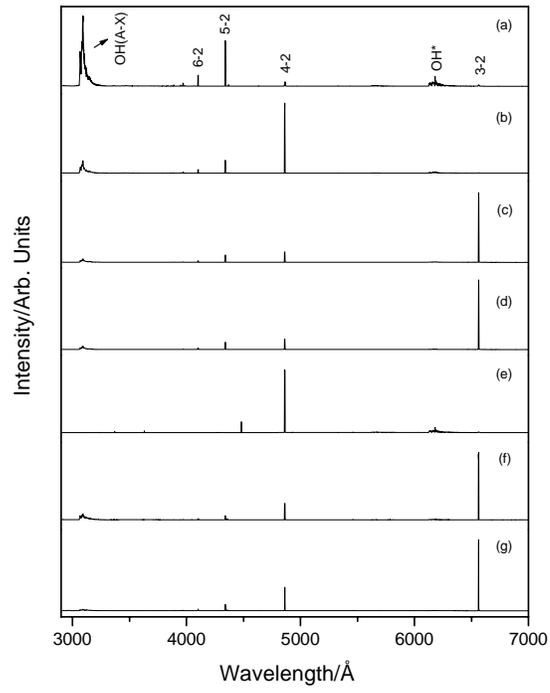

Figure 3

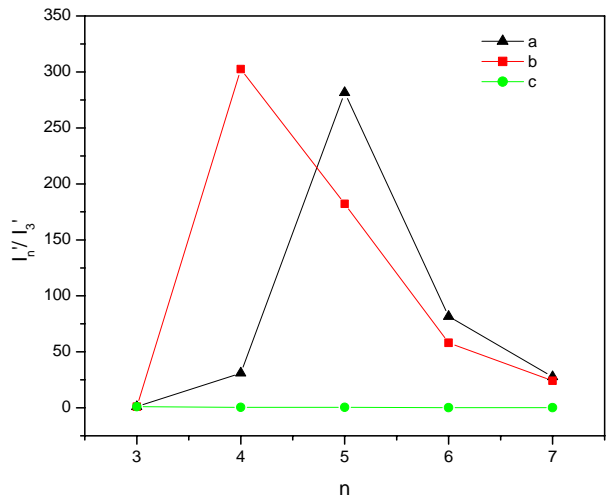

Figure 4

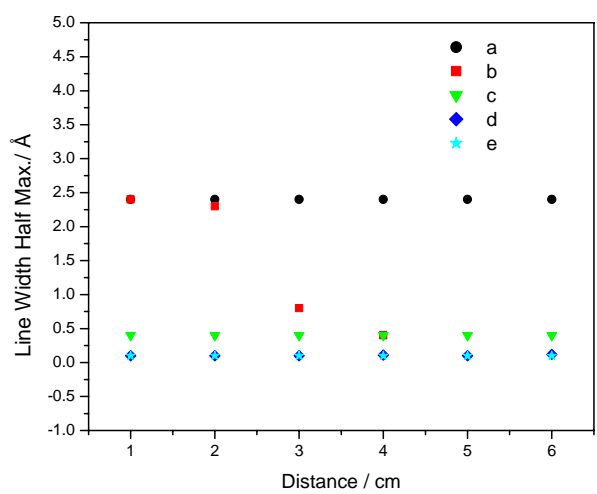

Figure 5

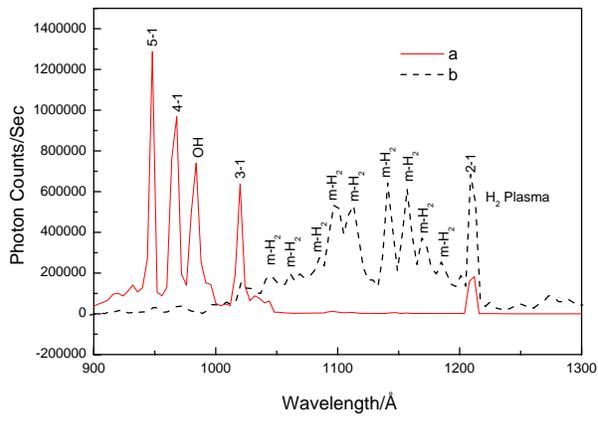

Figure 6

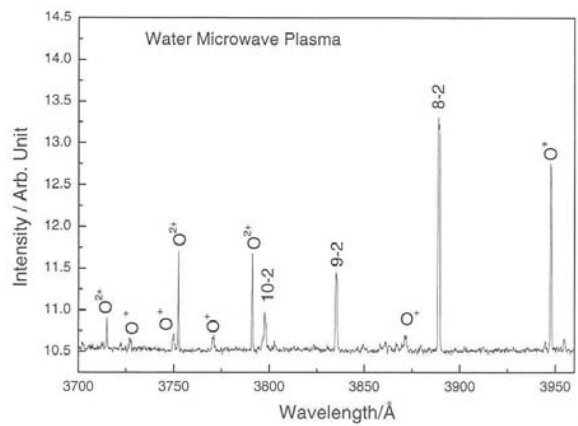

Figure 7